\magnification=\magstep1
\voffset=0.1truecm
\vsize=23.0truecm
\hsize=16.25truecm
\parskip=0.2truecm
\baselineskip=10pt

\def\np{\vfill\eject}
\def\teff{T_{\rm eff}}
\def\nth{n_{\rm th}}

\def\dmin{d_{\rm min}}
\def\sigbar{ {\langle \sigma \rangle}}

\def\fun#1#2{\lower3.6pt\vbox{\baselineskip0pt\lineskip.9pt
  \ialign{$\mathsurround=0pt#1\hfil##\hfil$\crcr#2\crcr\sim\crcr}}}
\def\ie{{\it i.e.}, }
\def\eg{{\it e.g.}, }
\def\etal{{\it et al. }}

\def\lglsun {\log (L/L_\odot)}

\def\dmax{d_{\rm max}}
\def\dmin{d_{\rm min}}
\def\mbol{M_{\rm bol}}

\def\sol{$_\odot$ }                     

\def\rfnce{\par\noindent\hangindent 20pt {}}
%
%
%
\bigskip
\bigskip
\centerline{{\bf MACHOs, White Dwarfs, and the Age of the Universe}}
\bigskip
\bigskip
\centerline{\bf David S. Graff$^{1,2}$, Gregory Laughlin$^1$, and Katherine Freese$^1$} 
\bigskip
\bigskip
\centerline{\sl $^1$Physics Department, University of Michigan}
\centerline{\sl Ann Arbor, MI 48109, USA}
\medskip
\centerline{\sl $^2$DAPNIA/SPP, CEA Saclay}
\centerline{\sl F-91191 Gif-sur-Yvette CEDEX, France}
\medskip
\centerline{graff@hep.saclay.cea.fr,
gpl@boris.physics.lsa.umich.edu, ktfreese@umich.edu}
\bigskip
\centerline{\it submitted to the Astrophysical Journal} 
\bigskip 
\centerline{\it 4 March 1997}
\bigskip
\bigskip
\centerline{\bf Abstract}
\vskip 0.1 in

A favored interpretation of
recent microlensing measurements towards the Large Magellanic Cloud
implies that a large fraction (i.e. 10--50\%) of the mass of the galactic
dark halo is composed of white dwarfs. However,
a ground based search by Liebert, Dahn \& Monet (1988; LDM) and a
recent search of the Hubble Deep Field by Flynn, Gould \& Bahcall(1996)
did not detect a substantial dark halo population of white dwarfs;
thus the putative halo population is either dim enough or sparse
enough to have eluded detection. In this paper we compare model white dwarf 
luminosity functions to the data from the observational surveys in order to
determine a lower bound on the age of any substantial white dwarf halo
population (and hence possibly on the age of the Universe).
In the course of our analysis,
we pay special attention to the velocity bias in 
the LDM survey; we show that (and quantify by how much) 
the velocity bias renders the survey significantly
less sensitive to a cool white dwarf population.

We show that the minimum age of a white dwarf halo population depends most
strongly on assumptions about three unknown quantities
({\bf 1}) the white dwarfs' total space density,
({\bf 2}) their atmospheric composition, and ({\bf 3}) the initial mass 
function (IMF) of their progenitors. We compare various
theoretical white dwarf luminosity functions, in which we vary these
three parameters, with the abovementioned survey results.
From this comparison, we conclude that if white dwarfs 
do indeed constitute more than 10\% (30\%)
of the local halo mass density, then the Universe {\it must} be at least
10 Gyr old (12 Gyr old) for our most extreme allowed values
of the parameters. When we use cooling curves 
that account for chemical fractionation
and more likely values of the IMF and the
bolometric correction, we find tighter limits:
a white dwarf MACHO fraction of 10\% (30\%) 
requires a minimum age of 14 Gyr (15.5 Gyr).
Our analysis also provides evidence that the halo white dwarfs
have helium-dominated atmospheres although this conclusion may change after
low temperature white dwarf atmospheres have been calculated.

\vskip 0.1 in\noindent
{\it Subject headings:} galaxies: halos -- stars: luminosity function,
mass function -- white dwarfs

\np
\centerline{\bf 1. INTRODUCTION}
\vskip 0.12 in

In this paper, we refine the process of using dark halo white dwarfs to
derive a lower bound on the age of the universe. Our efforts are
motivated by the second year MACHO collaboration data (Alcock \etal
1996) which appear to indicate that objects of white dwarf mass may
constitute a substantial fraction of the total mass of the galactic
dark halo.

If there is a profusion of white dwarfs in the galactic halo, one can
naturally consider the prospects for detecting them optically.  Kawaler (1996)
predicted that halo white dwarfs might be present in the Hubble Deep Field
(hereafter, HDF).  Flynn, Gould, and Bahcall 
(1996), Reid \etal (1996) and Elson, Santiago, and Gilmore (1996)
searched the HDF for this population (with rigorous selection criteria)
without notable success. Furthermore, the
ground-based white dwarf luminosity function compiled by Liebert,
Dahn, and Monet (1988; hereafter LDM) with stars from the Luyten Half-Second
Catalogue (Luyten 1979) shows
little evidence of a substantial population of halo white dwarfs.
The immediate conclusion is that if the white dwarfs exist, then they
are too dim to have been uncovered by the LDM and HDF surveys.
Our intention in this paper is to provide a detailed quantification of
this statement and its ramifications.

White dwarfs grow cooler and dimmer as they age. The cooling theory
for white dwarfs is well developed (e.g. Winget \etal 1987, Segretain
\etal 1994).  Cooling models, in conjunction with star formation
histories, initial mass functions (IMFs) and stellar evolution theory can be
used to construct model luminosity functions for posited white dwarf halo
populations.  The luminosity function is defined
to be the number density of stars with magnitude between
$M$ and $M+dM$.   By squaring the lack of observed faint white dwarfs in the
HDF and LDM studies with model luminosity functions, it is possible to derive a
lower bound on the age of the galactic halo, and 
hence on the age of the universe.

Unfortunately, 
the minimum age also depends on several loosely constrained parameters, 
notably the local density of halo white dwarfs, the present composition
of the white dwarf atmospheres, and the mass distribution of the white dwarfs
(which is in turn related to the IMF of the progenitor stars).
The primary results of this paper are plotted in Figures 4 and 5;
in obtaining these plots we have used the cooling
curves of Segretain \etal (1994) as communicated by G. Chabrier.
These figures indicate that if
one takes the local mass density in white dwarfs to be at least
10\% of the local halo mass density (as suggested by the microlensing
data), then the white dwarfs must be at least 12 Gyr old.
With the most reasonable mass
function for the progenitors, the lower limit is 13 Gyr;
if one also takes pure blackbody rather than the most extreme
bolometric corrections, the lower limit rises to 14 Gyr.  Although
relatively young, a universe of 12 Gyr is still in conflict 
with a Hubble constant $H_0 > 55$ km s$^{-1}$ Mpc$^{-1}$ in a 
standard cosmological model.  

Chemical fractionation provides an additional source of energy to
white dwarfs, and allows them to cool more slowly.  Models by Segretain et al which do not include C/O chemical
fractionation are about 2 Gyr younger than models with chemical fractionation.
With these simpler cooling models, all the age limits in the
previous paragraph become weaker by 2 Gyr.  Thus the extreme
minimum age one can accomodate for the white dwarfs is 10 Gyr.

If we assume that the age of the galactic halo is less than 
18 Gyr (Chaboyer \etal 1996), then we also find strong evidence 
that a substantial halo white dwarf population consists of white
dwarfs with Helium rather than Hydrogen atmospheres.  
The remaining uncertainty on this issue is due to the fact that
low temperature white dwarf atmospheres have not been studied,
although calculations by Mera and Allard are underway.

Please note that there is a potential source of confusion by our use of
the word ``halo''.  We distinguish between two different types of
galactic halos: the visible halo, sometimes called the stellar halo or
spheroid, which is dynamically insignificant in the outer Galaxy, and
the dark halo, or corona, which dominates the dynamics of the outer
Galaxy.  If much of the mass of the dark halo is composed of white
dwarfs, then it, along with the spheroid, is a stellar phenomenon;
however, the two stellar populations would have greatly different IMFs
and can thus be distinguished.  Microlensing events towards the LMC are
sensitive to a dark halo population (Alcock \etal 1997).  Thus, in this
paper, unless otherwise stated, we are exclusively concerned with a
dark halo population of white dwarfs.

\bigskip
\centerline{1.1 Comparison with Previous Work on this Subject}
\medskip
The white dwarf luminosity function of the galactic {\it disk} has been
extensively studied. The LDM survey (discussed in more detail below)
has been modeled by a number of groups, most significantly by
Winget \etal (1987). There is a fairly robust consensus that
the LDM survey is consistent with a disk population of white dwarfs
having an age of 9 -- 11 Gyr. A new survey of binary white dwarf companions
to main sequence stars, which is independent of the LDM survey, has
separately confirmed this result and yields a measurement of $t_{\rm
disk}\approx9.5$ Gyr (Oswalt \etal 1996).

Prior to our work, three groups constructed model
luminosity functions for {\it halo} populations of white
dwarfs, and compared them to the LDM luminosity function.
A prescient first investigation was made by Tamanaha {\it et al.} (1990),
who addressed the problem prior to the advent of microlensing surveys.
They recognized that a dark halo population of white dwarfs must
necessarily arise from an old (in their analysis older than 12 Gyr) population
of predominantly high mass (i.e. 2-6 M\sol ) stars. 
Motivated by the new MACHO results, Adams \& Laughlin (1996)
and Chabrier, Segretain, \&
M\'{e}ra (1996) revisited the problem and produced new
luminosity functions based on more sophisticated mass functions and
cooling curves. Their work reinforced the conclusion that any
significant white dwarf halo is the outcome of a very old population
of high mass progenitor stars. Adams and Laughlin (1996) obtained
a lower limit of 14 Gyr and Chabrier \etal of 18 Gyr for the
age of the white dwarfs.  Our work revisits this problem and
represents a substantial shift in that our results allow
a considerably younger minimum age for the white dwarf population.

This paper continues in the spirit of the three halo white dwarf
studies mentioned above.
We share the overall goal of comparing theoretical model luminosity functions
with observational surveys, and we have identified several specific areas
for improvement in the
comparison between observation and theory. Our aim is to address the
following issues and incorporate them into a more refined analysis. 
These improvements lie in three areas: 
1) the method of comparison with data,
2) the treatment of bolometric corrections,
and 3) comparison of theory with the combined results of the
HDF and LDM surveys.
In the course of this analysis,
we emphasize the importance of a little-publicized velocity bias
in the LDM survey.  Each of these three areas of improvement
is outlined below.

{\bf 1) The Method of Comparison:} 
Two of the groups who previously worked on this subject,
Tamanaha \etal (1990) and Adams \& Laughlin (1996),
focused on the dimmest white dwarfs observed by LDM, which
had a visual magnitude of $M_V \approx 16$.  These two groups compared
predictions from their luminosity functions with these observed
white dwarfs.  The third group, Chabrier \etal (1996), used 
the {\it nondetection} of white dwarfs by LDM at fainter magnitudes
to compare with theory.  We also use the nondetection,
but in a more systematic way.

{\it Comparison with the Coolest White Dwarfs Observed:}
Tamanaha \etal (1990) and Adams \& Laughlin (1996) attempted to constrain the 
age and number density of the dark halo population by focusing on the
dimmest white dwarfs which were actually uncovered in the LDM survey. The LDM
luminosity function exhibits a sharp dropoff in number density for objects
fainter than $M_V \approx 16$. Tamanaha \etal (1990) and Adams \& Laughlin 
(1996) required their models to conform to
the density of $M_V \sim 16$ stars from LDM data.
Comparisons of this type, namely of theory with the coolest
white dwarfs actually observed, have two shortcomings: they are sensitive
to small differences in the low mass tail of the progenitor IMF, and
they ignore the bulk of the posited halo white dwarfs, which
would be dimmer than the faintest stars detected by LDM. 
We briefly discuss these two shortcomings here.

The sensitivity to the low mass tail can be understood
as follows. 
In Fig. 1, the horizontal dotted line shows
the bolometric luminosities of the dimmest
white dwarfs observed by LDM (here we have assumed a blackbody bolometric
correction; see the discussion below).  
The solid lines plot model white dwarf luminosities.
We can see from Fig. 1 that for a white dwarf population older than
10 Gyr, only the stars with the lowest initial masses would
be bright enough to be limited by the LDM luminosity function 
at $M_V = 16$.
 
White
dwarfs which emerged from progenitors of $M< 1.5 M_{\odot}$ are substantially 
brighter than white dwarfs which arose from heavier stars. 
[The lower mass progenitor stars were on the main
sequence for a significant portion of a Hubble time, and their
resulting low mass white dwarfs also cool more slowly.]
The brightest prospective
halo dwarfs are strictly a product of the tail of the 
progenitor IMF, which can be subject to a fair amount of variation without 
radically changing the underlying star formation theory. Hence
the density of low mass progenitors and resulting brightest
white dwarfs is not well known. 
In addition, a focus in the method of comparison 
on the faintest detected stars tends to ignore the bulk of the 
posited halo white dwarfs, as it does not aggressively take into account the
absence of observed stars at low luminosity.

{\it Comparison of Theory with Nondetection of Very Cool White Dwarfs:}
In contrast to the situation in the disk,
our knowledge of the halo white dwarf
luminosity function comes primarily from an {\it absence} of
observed stars. The most restrictive
comparison with theoretical models is thus a subtle
matter of finding limits on a posited population, rather
than the more straightforward task of
determining the best model fit to a set of data points. 
At best, we can only delineate the allowed regions of 
a parameter space which encompasses the age and halo mass
fraction of the lensing population.  
By using the {\it nondetection}
of faint white dwarfs in various data sets, one can avoid
undue emphasis on the low mass tail of the IMF and can take advantage
of the absence of observed stars at low luminosity.

Chabrier \etal (1996) avoided an undue emphasis on the tail of the IMF
by taking advantage of a particular interpretation of LDM's {\it
nondetection} of faint white dwarfs and using it to constrain their
model luminosity functions. Although LDM did not detect any stars
indicative of a very faint halo white dwarf population, they could
have if the white dwarfs had been present in sufficient
quantity. Indeed, LDM attempted to quantify this possibility by
indicating the ficticious luminosity function which would have
resulted had a single white dwarf been found in each of their
unoccupied faint bins below $M_V=16.$ Chabrier \etal (1996) used this
illustrative luminosity function to constrain the space density of
white dwarfs fainter than $M_{V}=16$. Although qualitatively
reasonable, this method is not the best way to obtain an upper limit
to the local white dwarf density.  Requiring that model luminosity
functions fall on or below the envelope of ficticious points does not
provide a uniform constraint. Among the models presented by Chabrier
\etal (1996), luminosity functions predicting nearly five stars among
the dim bins are no more constrained than others in the same figure
which predict only $\approx 1$.  In addition, we will show that the
LDM sensitivity to disk stars used by Chabrier \etal is up to
$10\times$ too restrictive for a population of halo stars (see the discussion
of proper motion sensitivity of the survey in section 3.2); this fact
led Chabrier \etal to overestimate the minimum age of the halo white
dwarfs.


In this paper, we also use the nondetection 
of faint white dwarfs in the LDM survey with $M_V > 16.5$
to constrain the properties of prospective halo populations. 
However, we take a more rigorous statistical approach:
we integrate over the entire low luminosity region of the luminosity
function to predict the number of stars which would be observed on average.
With the assumption that the white dwarfs are spatially uncorrelated,
Poisson statistics indicate that the number of dwarfs 
predicted by a given model luminosity function
must be ${{\cal N}\le 3}$ in order to be consistent with a nondetection at the
95\% confidence level. This statistical approach
allows a systematic test for consistency, which can be
applied automatically to a large number of trial model luminosity functions
to efficiently determine the allowed region of parameter space.

{\bf 2) Bolometric Corrections:}
Bolometric corrections are required in order to transform a white dwarf's 
observed $V$, $R$, or $I$ band magnitude into a bolometric luminosity;
it is this latter quantity that is predicted by theoretical models.
Tamahana {\it et al.}, Chabrier {\it et al.} (1996), and Adams \& Laughlin
(1996) compared their results to the {\it bolometric} luminosity
function of LDM, and thus made implicit use of bolometric corrections
introduced by LDM.  While fairly accurate for the brighter white dwarfs
comprising the disk population, these bolometric corrections cannot
be used to extrapolate to visual magnitudes fainter than $M_{V}$=16
and thus do not apply to the presumed halo  white dwarf population.  We
apply a more sophisticated estimate of the bolometric corrections appropriate
to extremely dim white dwarfs. We also focus discussion on the possibility
that the putative halo dwarfs have hydrogen-dominated atmospheres. Hydrogen
atmospheres imply very different bolometric corrections, as well as 
considerably
longer cooling times. We show that a preponderance of hydrogen atmospheres
is extremely unlikely in the event that the halo is younger than 18 Gyr.

{\bf 3) Consolidation of Surveys:} 
In addition to the LDM survey used by previous authors to compare
with theoretical luminosity functions, we also examine the search of
the Hubble Deep Field (HDF) by Flynn, Gould, and Bahcall (1996).
The HDF and the LDM survey can be used independently to
place constraints on the age and number density of a halo white dwarf
population.
We put forth a detailed analysis of the effective volume
searched by each of these surveys as a function of bolometric
luminosity, which allows us to use the combined results of both surveys to place
firmer limits than possible with either survey alone. In the course
of our analysis, we discuss a little-publicized velocity bias
in the LDM survey which renders it less sensitive to the
faint, high proper motion stars which are characteristic of
the population we are studying.  In fact, taking the limit properly
into account will admit considerably younger halo ages than found in prior
analyses.

The organization of this paper is as follows: In \S 2, we describe the
theoretical issues involved in the construction of model white dwarf
luminosity functions. These center mainly on the cooling theory and the
star formation theory endemic to the original population. In \S 3, we
examine issues related to the comparison between observations and theory.
We characterize the surveys, take stock of bolometric corrections, and 
discuss the question of faint white dwarf atmospheric composition. In \S 4, we 
compare our theoretical luminosity functions with observations to
constrain the age and number density of the putative halo population. In \S 5, 
we discuss the ramifications of our work within a larger cosmological
framework.
\bigskip 
\centerline{\bf 2. MODEL LUMINOSITY FUNCTIONS}
\medskip
The white dwarf luminosity function (LF) 
describes the number density of stars
per magnitude interval.
We need to obtain a theoretical LF for the white dwarfs
in order to compare with the LDM and HDF data.      
White dwarf luminosity functions hinge on the physical properties of their
component white dwarfs. The luminosity of a particular white dwarf is
determined by its age, its mass, and its chemical composition. 
These properties in turn depend
on the properties of the main sequence progenitor star that
led to the white dwarf.

We will assume that all the main sequence progenitors of the
present lensing population formed at a time $t_{\rm H}$ ago.
Then we have $$t_{\rm evol}(m)+t_{c}(l,m_{\rm WD})=t_{\rm H}. \eqno(2.1)$$
Here, $t_{\rm evol}(m)$ is the time the progenitor spent on the main 
sequence, and $t_c$ is the time
the star has spent as a white dwarf.
We assume that the nuclear burning lifetime on
the main sequence is a function
only of the mass $m$ of the main sequence progenitor, i.e.,
$$t_{\rm evol} \equiv t_{\rm evol}(m) \, . \eqno(2.2)$$
We will also use the fact that the mass of the white dwarf is a
function of the mass of its main sequence progenitor,
$$m_{\rm WD} \equiv m_{\rm WD}(m) \, . \eqno(2.3)$$
Note that lower mass stars spend a longer time on the main sequence,
become white dwarfs later, and hence are brighter at any
given time (see Figure 1).

White dwarf progenitor lifetimes are
assumed to conform to the relation,
$$\log_{10} t_{\rm evol} = 9.921 - 3.6648 (\log_{10} m) +
1.9697(\log_{10} m)^{2} - 0.9369(\log_{10} m)^{3} \, , \eqno (2.4)$$
where $m$ is measured in units of solar mass.  
This result is taken from Iben \& Laughlin (1989) who
obtained the polynomial by extracting main sequence lifetimes from the
stellar evolution calculations of a number of different authors.

The cooling time of the white dwarf is a function of the white
dwarf mass and of the particular luminosity interval $l$.
We will assume
that separate relations of the form
$$t_{c} \equiv t_{c}(l,m_{\rm WD}) \eqno(2.5)$$
hold for white dwarfs with hydrogen and helium atmospheres.
We primarily use the white dwarf cooling theory 
from the calculations of Segretain \etal (1994), as communicated
by G. Chabrier (1996). The Segretain
\etal (1994) model represents an advance over earlier cooling theories
in that it accounts for gravitational energy release
due to carbon-oxygen differentiation at crystallization. Proper treatment
of crystallization yields significantly longer white dwarf cooling times, which
in turn imply an older age for any  particular white dwarf halo population.
These white dwarf models correspond to a mass sequence of initially
unstratified white dwarfs composed of equal parts carbon and oxygen, with
helium atmospheres.  

To estimate the uncertainty
in the age limits for white dwarfs due to cooling theory,
we note that an extreme model presented by Segretain \etal which does
not include the perhaps controversial Carbon-Oxygen phase separation
and which allows an initial C/O stratification predicts that cool
white dwarfs would be $\sim 1$ Gyr younger than the homogeneous
phase separated models we use.  These models
would give rise to minimum white dwarf ages 2 Gyr
younger than what is allowed by the 
Segretain (1994) models, and hence potentially allow
a halo white dwarf age of 10 Gyr.

If we assume that all of the white dwarfs were formed within
a relatively short period of time, then one can
use equations (2.1) and (2.3) to derive a differential form
for the luminosity function (see Iben \& Laughlin 1989
for details),
$$ \Phi(\mbol) \equiv {dn\over{dM_{\rm bol}}}=
{ 0.921n_{\rm WD}\, (dN/dm) \, \, t_{c}\, \, (\partial \log_{10} t_{c}
/\partial \log_{10} l)_{m_{\rm WD}} \over
{(dt_{\rm evol}/dm) + (t_{c}/m_{\rm WD})
(\partial \log_{10} t_{c} / \partial \log_{10} m_{\rm WD})_{l}
(d m_{\rm WD} / dm)}} \, . \eqno(2.6)$$

{\it MACHO fraction:}
We will normalize Eq. (2.6) to give a white dwarf
number density appropriate to a given MACHO fraction of the local
halo density.
The quantity $n_{\rm WD}$ in equation (2.6) is the local number density of
halo white dwarfs.  Guided by the microlensing results, we
will hereafter equate this $n_{\rm WD}$ with the local number density
of MACHOs.  One must always bear in mind that this assumption is
not necessarily true; \ie the MACHOS are not necessarily
white dwarfs.  
This number density can be converted into
a mass density $\rho_{\rm WD}$
if one multiplies by the average white dwarf mass;
this average white dwarf mass can be obtained from a given initial
mass function for the progenitors of the white dwarf together
with an initial/final mass relation described below.  For all cases, 
the average mass is near 0.6$M_\odot$, consistent with the best fit mass
extracted from the MACHO data.

We then define the quantity $f$ to be the MACHO fraction
of the halo density, i.e., the ratio of the mass density of MACHOs
over the local mass density of the halo.
We have taken the local density of the entire halo
to be $10^{-2} M_\odot/pc^3 $ (\eg Bahcall and Soneira 1980);
this number of course depends on the galactic model chosen.
The MACHO fraction in turn is obtained from the observed optical depth
of the lensing population in conjunction with a particular model of
the baryonic mass distribution of the halo. Alcock \etal (1996) have
shown that although the derived MACHO mass fraction depends strongly on the
choice of halo model, the MACHO fraction exceeds 0.1 with at least 90\% 
confidence for every model they considered. Turner, Gates \& Gyuk (1996) have 
independently determined that the vast majority of otherwise plausible halo
models must have $f>0.1$ in order to be consistent with the microlensing
result. As our working hypothesis we therefore adopt $f>0.1$ 
as the minimum value in our analysis.

{\it Initial Mass Function:} As mentioned above, the mass and main
sequence evolution of a white dwarf are determined by the mass of its
main sequence progenitor.  Therefore we must examine possible initial
mass functions (IMFs), $dN/dm$, for main sequence stars that are
capable of producing white dwarfs.  Optical searches are
more sensitive to brighter white dwarfs arising from low mass
progenitors than to the fainter dwarfs arising from high mass
progenitors; thus the limits that we will be able to
place depend on the luminosity function we choose.  We compute a wide
range of luminosity function models to measure the sensitivity of our
limits on the luminosity function.
	
AL argued that the initial masses of halo white dwarf progenitors had to
be between 1 and 8 M$_\odot$.  The lower limit on the range of
initial masses comes from the fact that stars with mass $< 1 M_\odot$
would still be on the main sequence.
The upper bound arises from
the fact that progenitor stars
heavier than $\sim 8 M_\odot$ explode as supernovae, and do not form
white dwarfs.

Because low mass main sequence halo stars are intrinsically scarce
(Bahcall \etal 1995; Graff \& Freese 1996a,b),
an IMF of the usual Salpeter (1955)
type $dN/dm \propto m^{-2.35}$
is not appropriate, as it would imply a gross overabundance of low mass
stars in the halo. We follow Adams \& Laughlin (1996) and use a log-normal
mass function motivated by Adams \& Fatuzzo's (1996) theory of the IMF:
$$\ln {dN \over dm}(\ln m) = A - {1 \over 2 \sigbar^2}
\Bigl\{ \ln \bigl[ m / m_C \bigr] \Bigr\}^2 \, . \eqno(2.7)$$
The parameter $A$ sets the overall normalization. The mass scale
$m_C$ (which determines the center of the distribution) and the
effective width $\sigbar$ of the distribution are set by the
star-forming conditions which gave rise to the present day population of
remnants. For our standard case, we take the parameters $m_C=2.3 M_{\odot}$
and $\sigbar=0.44$, which imply warm, uniform star-forming conditions
for the progenitor population. These parameters saturate the twin constraints
required by the low-mass and high-mass tails of the IMF, as discussed
by Adams \& Laughlin (1996), i.e., this IMF is as wide as possible. 

The IMF in Eq. (2.7) is different from the mass functions
measured in the disk, the halo, globular clusters, and elliptical galaxies, and
justifiably strikes many as being a sign of fine tuning required for a white
dwarf halo model.  Indeed a MACHO IMF must be different
from the standard IMFs and must be narrow. It is interesting to note,
however, that the star formation theory
of Adams and Fatuzzo (1996) predicts that a zero metallicity
primordial gas would form higher mass stars than a non zero
metallicity gas which formed all the familiar stars; this prediction lends 
some plausibility to Eq. (2.7).

As an even more extreme possibility, we
will also explore the effects of
various single valued (delta function) IMFs, mainly for illustrative
purposes.  Such a delta function mass function is undoubtedly too
narrow and unphysical.  However, it allows us to eliminate the poorly
constrained low mass tail of the IMF, which produces the most easily
observed high luminosity tail of the LF.  Thus, delta function
mass functions enable us to focus on the effects of the initial
mass of the majority of stars, without having to worry about the shape of the
wings of the IMF.


{\it Initial/Final Mass Relation:}
The relation between the mass of a progenitor star and 
the mass of its resultant white dwarf is rendered uncertain by an
imperfect understanding of mass loss from red giants.  We follow Wood
(1992) and adopt his standard transformation between the progenitor
mass and the white dwarf mass,
$$m_{\rm WD} = A_X \exp [ B_X m ] \, , \eqno(2.9)$$
with $A_X$ = 0.49 and $B_X$ = 0.095. 
With our choice of IMF in Eq. (2.7), 
this leads to an average White Dwarf mass of
0.63 M$_{\odot}$. Adopting other reasonable Initial/Final relations,
such as the form $m_{\rm WD}=0.45+0.1m$ given by Iben \& Tutukov (1984),
has little qualitative effect on the results.
We estimate that the uncertainty in the initial/final mass relation
imparts an uncertainty of less than 0.5 Gyr to our final answer
of the age of the white dwarf population.

{\it Atmospheric Composition:}
Tamanaha \etal (1990), Adams \& Laughlin (1996), and Chabrier \etal (1996),
all assumed that the putative white dwarf halo population is 
composed of dwarfs with helium atmospheres. Faint white dwarfs with helium
atmospheres cool more quickly than dwarfs with hydrogen atmospheres
(due to lower helium opacities below 10,000K), and so any lower age limits
derived with the assumption of helium atmospheres are firm.
Nevertheless, hydrogen atmospheres are a possibility.

The chemical composition of cool white dwarf atmospheres is difficult to
determine observationally, depends on the metallicity of the
star (Iben and MacDonald 1986), and can change over the course of a white
dwarf's lifetime (Wesemael and Fontaine 1987).
Until recently, it seemed that the most likely 
case for very cool white dwarfs in the {\it disk}
is that they have Helium atmospheres.  Observations of LDM
indicated that this is probable, and Fontaine \& Wesemael (1987) suggested that
Hydrogen atmospheres could be diluted with the Helium mantle through convection.
However, a recent survey of the atmospheric type of cool white dwarfs
(Bergeron, Ruiz \& Leggett 1997)
found a substantial population of cool $(4000<T<6000)$ hydrogen atmosphere
stars.
These authors proposed a complex
interaction of convection and accretion from the interstellar
medium to allow the atmospheres of white dwarfs to rapidly
evolve from Helium to Hydrogen and back to Helium.
It is an interesting question to investigate the atmospheric
properties of even cooler white dwarfs, including the ones
we are considering as a halo population.
The atmospheric
structure of halo white dwarfs cooler than 4000K has never been observed.

The composition of white dwarf atmospheres is not
yet certain even in the disk, and certainly is unknown in the Halo.
The halo white dwarfs are likely to 
come from Pop III, zero metallicity stars,
with a very different IMF than that which produced
Pop II stars, and are much dimmer and cooler than 
any disk population of white dwarfs.  They are likely to accrete
much less hydrogen (from the disk interstellar medium).
In addition, hydrogen which is pressure ionized at low temperatures may
not effect the total opacity, and may thus be invisible (Bergeron, Ruiz
\& Leggett 1997).

The metallicity can play an important
role in determining the atmospheric composition, as illustrated
by Iben and MacDonald (1986) for Pop II stars.
The halo white dwarfs are older, have lower metallicity precursors,
and come from a different IMF than the disk white dwarfs. 
We feel that it is necessary to
investigate both Hydrogen and Helium atmospheres
for the dim halo white dwarfs.  Our most important result
will be our lower limit for the age of pure Helium atmosphere
white dwarfs, since these white dwarfs cool the most slowly
and hence correspond to the lowest possible ages.

Oswalt \etal (1996) have recently illustrated that if the 
observed {\it disk} population of faint
white dwarfs possesses predominantly hydrogen atmospheres, then the oldest
observed disk dwarfs are several billion years older than currently assumed 
(i.e. $t_{\rm disk}\approx 13$ Gyr as opposed to $t_{\rm disk}\approx 10$ Gyr).
Differences in bolometric corrections appropriate to the two atmospheric
classes will exacerbate this age effect at the
very low temperatures endemic to halo dwarfs fainter than $M_{V}\approx$16.
In order to explore the broad effect of hydrogen-dominated
atmospheres, we adopt the approximation that faint white dwarfs
containing a thin hydrogen atmosphere above a helium mantle cool 
30\% more slowly than the Segretain \etal (1994) models which have pure
helium atmospheres (M. Wood, personal communication).


\bigskip
\centerline {\bf 3. THE INTERPRETATION OF OPTICAL OBSERVATIONAL SURVEYS}
\medskip
In this section we discuss in detail how the LDM and HDF
surveys can be used to constrain the true nature of the lensing
population. In order to properly reconcile these two data sets with
the MACHO data, several subtleties must be taken into account. These 
include bolometric corrections to observed $V$, $R$, and $I$ band magnitudes,
and the oft-neglected proper motion cutoff present in the LDM survey.
\bigskip
\centerline { 3.1 Bolometric Corrections}
\medskip
Bolometric corrections provide a basis of comparison between model
luminosity functions, most naturally expressed as bolometric (total)
luminosity ($\mbol$), and observations made through specific bandpass filters.
The HDF exposure was made in the $I$ band; the original
Luyten Half-Second Catalogue (on which the LDM study is partially based)
had an $R$ band magnitude limit; and LDM's follow-up photometry was done
in the $V$ band. Bolometric corrections are therefore necessary for our
analysis. 

With the usual definition,
\footnote{$^{*}$}{All BCs were defined so that $\mbol(\odot)=4.72$.}
$$M_{\{I,R,V\}}=\mbol-{\rm BC}_{\{I,R,V\}}, \eqno(3.1)$$
a negative bolometric correction implies a larger value for $M_{\{I,R,V\}}$
and hence an object which appears less luminous in the frequency
band of interest. A conservative analysis
(i.e. an analysis which obtains a 
conservative lower limit on white dwarf ages)
should therefore adopt the most negative estimate of a
particular bolometric correction.

LDM adopted two extreme bolometric corrections to bracket the range of
possibilities for their dimmest white dwarfs (which they assumed had
helium atmospheres). For a subset of their stars, they had measurements
of both $M_{V}$ and $\teff$. To produce a lower bound on the relevant
bolometric correction, they assumed that the white dwarfs in this subset
were blackbodies, and then used the known values of $\teff$ and $M_{V}$ to
derive bolometric magnitudes. From the resulting plot of $\mbol$ vs.
$M_{V}$, they obtained the following relation:
$$\mbol=-19.55+3.847M_{V}-0.1042M_{V}^{2}. \eqno(3.2)$$
This fit is accurate for the range of magnitudes exhibited by the
stars in their sample, but it cannot be extended beyond $\mbol=16$
(the solution for $M_V$ in Eq. (3.2) becomes imaginary).
As an upper bound to the bolometric correction for helium atmospheres,
LDM adopted $\mbol=M_{V}$; that is, they took a zero value for the 
bolometric correction in this case.

Bergeron, Saumon, \& Wesemael (1995) have recently improved the estimation
of bolometric corrections by computing new hydrogen
and helium-dominated white dwarf model atmospheres. Unfortunately 
their models extend down to temperatures of only 4000K, whereas the bulk 
of the lensing population of halo dwarfs should have temperatures 
ranging between 2000 and 4000K. We thus need to make conservative 
estimates of the bolometric corrections in this low temperature regime
in order to help put a firm lower bound on the age of the halo population.
\bigskip
\centerline {3.1.1 Bolometric Corrections for Helium Atmospheres}
\medskip
The opacity in cool helium white dwarf
atmospheres is dominated by He$^{-}$ absorption.
However, in  a zero metallicity, pure He atmosphere,
the free electrons required to make He$^{-}$
are scarce.  In the limit of a vanishing He$^{-}$
fraction, the next largest opacity source is very weak Rayleigh scattering
(Bergeron, Saumon \& Wesemael 1995). The essential point is that cool
helium atmospheres have very low opacity, which causes white dwarfs with
pure helium atmospheres to cool relatively quickly.

Since He$^{-}$ opacity is largely frequency independent, one expects that,
to first approximation, the spectrum of a pure helium atmosphere will resemble
a blackbody distribution. Therefore, for pure helium atmospheres with T$<$4000K,
we adopt black body bolometric corrections (Allard 1990) with a multiplicative
``safety'' factor of 0.8,
$${\rm BC_{He}}(\teff)={\rm BC_{bb}}(0.8\teff). \eqno(3.3)$$
The multiplicative factor ensures that we are underestimating the bolometric
correction, and placing a conservative lower limit on the halo age.  
We will also make calculations for black body bolometric corrections 
(without the correction factor of 0.8), as these
are more likely to be representative of a helium 
atmosphere.  The differences in our results will illustrate the
range of uncertainty due to the fact that there are as yet no calculations
of model atmospheres for such cool white dwarfs.

\bigskip
\centerline {3.1.2 Bolometric Corrections for Hydrogen Atmospheres}
\medskip
Saumon \etal (1994) have published models of zero-metallicity, {\it red 
dwarf} atmospheres which span the (2000-4000)K temperature range. In the Saumon
\etal (1994) models, strong H$_{2}$ absorption in the infrared forces the bulk
of the radiation to emerge in the $R$ and $I$ bands. 
The far infra-red absorption becomes more significant for cooler
atmospheres as more H$_2$ is formed and as the blackbody peak
moves into the far-infrared, with the odd result that, as the
temperature drops, the peak of the emission spectrum moves
to {\it higher} frequencies.  Very dim zero-metallicity red dwarfs
therefore radiate quite efficiently in the $R$ and $I$ bands. 
In the Saumon \etal models, as the surface gravity is raised 
by two orders of magnitude, the atmospheres radiate more and more 
efficiently in the $R$ and $I$ bands, and become less like black body 
atmospheres.  Therefore, we expect Hydrogen atmosphere
white dwarfs, which have surface gravities approximately three orders
of magnitude larger than the highest gravity Saumon \etal models,
to emit an even
higher proportion of their radiation in the $I$ and $R$ bands,
exactly the bands of the observations. We thus
choose to adopt the bolometric corrections appropriate to the Saumon \etal
(1994) red dwarf atmospheres as a lower limit to the corrections appropriate
to high surface gravity white dwarfs. As shown in Figure 2 (in which
all of our bolometric correction estimates are plotted) red dwarf bolometric
corrections are less than the white dwarf bolometric corrections in the
region of overlap, $4000K < \teff < 5000K$, and it is encouraging that the 
red dwarf and hydrogen white dwarf atmospheres show a similar trend of behavior
in this regime.

Since white dwarf atmospheres have a surface gravity three orders of
magnitude higher than the red dwarf atmospheres we use, it is possible
that new physics will govern the white dwarf Hydrogen atmospheres.  In this
case, we would expect their bolometric corrections to be very
different from the red dwarf bolometric corrections that we have
adopted.  We reserve the right to alter our conclusions once
white dwarf atmospheres of this low temperature have been
calculated.  We do wish to emphasize, however, that the lower
age limits obtained from the pure {\it Helium} atmosphere white
dwarfs discussed in the previous section are firm; 
our estimates of bolometric corrections for the case 
of Helium atmospheres discussed in the previous section are
extremely conservative and the age limits for Helium atmosphere
white dwarfs are expected to become
more constrained in the future.

White dwarfs with hydrogen-dominated atmospheres are far more detectable than 
dwarfs with helium atmospheres. The strong H$_2$ absorption keeps the
opacity of the Hydrogen atmosphere white dwarfs relatively high and thus forces
them to cool relatively slowly.  Thus hydrogen dwarfs are inherently
brighter.  In addition, as discussed above,
they also emit their radiation preferentially in the frequency
bands to which the surveys are sensitive. Since hydrogen-atmosphere
dwarfs are so much easier to see, the limit that we can place on the
number density of these stars
is considerably more severe.
In order to be in agreement with the observed paucity of white dwarfs
in the HDF and LDM data, hydrogen dwarfs would have to be {\it very}
old, as discussed below.  In fact the brightness difference between
Hydrogen and Helium white dwarfs leads to an astounding
difference of more than 8 Gyr in the inferred age of a population.
\bigskip
\centerline {3.2 The Effective Volume of the LDM Luminosity Function}
\medskip
LDM culled their luminosity function from a survey of high proper
motion stars carried out by Luyten (1979). The Luyten survey was sensitive
to stars with $R$ magnitudes brighter than 18 and
proper motions in the range  $0.8''<\mu<2.5''$ (Liebert \etal 1979).
Due to a need to minimize computer time, very high proper motion stars
with $\mu>2.5''$ were {\it not} catalogued. 

The velocity bias present in the Luyten survey has little consequence for
the luminosity function of disk white dwarfs.
As a result, it has been scantily documented.
For a high velocity halo population, however, it is significant. The proper motion 
cutoff in Luyten's survey implies that the tangential velocity of a 
white dwarf detected by the survey must be 
$$V_{\rm tan}<121 {\rm {km \over sec}}\,(d/10{\rm pc}). \eqno(3.4)$$
Thus a white dwarf with a typical halo tangential velocity of 
270 km/sec must be at least 22 parsecs away to have been registered in the
survey. This distance corresponds to $M_R < 16.4$,
since the maximum distance a star with absolute magnitude
$M_R$ can be seen is $\dmax = 10^{[(18-M_R)/5+1]}$pc.
If we assume blackbody atmospheres, we have very roughly
$(V-R) \sim 1$ so that the corresponding visual magnitude
is $M_V < 17$.  Thus the velocity bias against high velocity stars in the LDM
survey becomes significant at $M_V \sim 17$, while some of the white
dwarfs predicted by the Adams and Laughlin luminosity
function have magnitudes down to $M_V \sim 18$.
Due to the upper limit on the proper motions,
a significant number 
of the white dwarfs predicted by the previously described model
luminosity functions would have been missed by the LDM survey.

To assess the degree of incompleteness intoduced by the proper motion
cutoff, we have calculated the effective volume probed by the Luyten
survey. Effective volume is defined to indicate the actual volume
probed weighted by the probablity that a halo star would have a
sufficiently low proper motion to appear in the survey, {\it i.e.},
$$V_{\rm LDM}^{\rm eff}=\int\limits_{\vec{x}\in{\rm \{vol \}}}
d^3\vec{x} \,  P(\vec{x}) \, .
\eqno(3.5)$$
where $\{{\rm vol}\}$ is the northern third of the sky, to a distance 
$\dmax$ which is a function of luminosity. $P(\vec{x})$ is the probability
that a halo star at location $\vec{x}$ would have the appropriate proper
motion,
$$P(\vec{x})=\int\limits_{0.8<\mu(\vec{x},\vec{v})<2.5} d^3\vec{v} \,
Q(\vec{v})\, . \eqno(3.6)$$
 Here $Q(\vec{v})$ is the probability density
that a halo star has velocity $\vec{v}$. We assumed that the local
halo stars have an isotropic Maxwellian velocity distribution with a
dispersion $\sigma_{v}=270$km/sec, and that the sun moves through this
distribution with a velocity $v_{\odot} =220$ km/sec.
$\mu(\vec{x},\vec{v})$ is the proper motion, $v_\perp/|x|$.

Figure 3 shows the effective volume of the LDM survey
as a function of white dwarf bolometric
luminosity, as well as the volume which would have been searched had the full 
range of proper motions been tested, i.e. 
$V_{\rm LDM}^{\rm max}=(\Omega/3) \dmax^3$. (The volume searched by the
HDF survey is also shown, and is discussed in the next section.) 
The Luyten survey would have been substantially more sensitive to halo
dwarfs had there been no {\it ad hoc} proper motion limit. Taking
the limit properly into account will admit considerably younger halo ages than 
found in prior analyses. 

As mentioned previously, the result of the LDM survey that
is relevant to this work is that no white dwarfs were found
dimmer than $M_V = 16$.  Below we will square this result 
with the MACHO experiment's suggestion that the halo is full of white
dwarfs.

\bigskip
\centerline {3.3 Effective Volume of the Hubble Deep Field}
\medskip
The Hubble Deep Field exposure taken by Williams \etal (1996) is sensitive
to white dwarfs which are much more distant than those in the Luyten survey.
Flynn \etal (1996), Reid \etal (1996) and Santiago \etal (1996) examined the HDF with the goal
of determining an upper limit on the local
density of white dwarfs.  In this paper, we will focus on the analysis of
Flynn \etal.  The other authors derived similar negative results which would
allow us to place similar limits.  However, the fact that three
independent teams derived similar results suggests that the dearth of
white dwarfs in the HDF was not due to problems in the analysis of one
of the groups.

The analysis of Flynn \etal was sensitive to stars
in the magnitude range $24.6<I<26.3$,
where the bright cutoff was made to
avoid a foreground contamination of the sample by disk stars.
(The discarded volume constituted only 10\% of the total sample volume.)
Within their sample, Flynn {\it et al.} (1996) found
one stellar object with ($V-I)=1.32$ and $I$=23.71. 
They saw none which were either redder 
or dimmer. If the single detected star-like object is a white dwarf with
a helium atmosphere, the atmospheric models of Bergeron, Saumon \& Wesemael (1995)
indicate an effective temperature of 4050K. For a $0.6 M_{\odot}$
white dwarf, this corresponds to a luminosity of $\lglsun=-4.0$.

Although the HDF probes far deeper than the Luyten survey,
its 4.4 square arcminute field of view is much smaller. Employing
both hydrogen and helium $I$-band bolometric corrections, we computed the maximum and minimum distances surveyed as
$$d_{\rm max}^{\rm min} = 10^{0.2(M_I-I^{\rm max}_{\rm
min})+1}{\rm pc}, \eqno(3.7)$$
and calculated the effective volume of the HDF search as
$$V_{\rm HDF}=(\Omega/3) (\dmax^3-\dmin^3), \eqno(3.8)$$
for a range of white dwarf bolometric luminosities.
The results are shown in Figure 3.

As mentioned above, the surveys are much more sensitive to
white dwarfs with hydrogen-dominated atmospheres.
For hydrogen atmospheres, the
LDM survey is always much more sensitive than the HDF.
For the more likely case of
helium atmospheres, the LDM survey is more
sensitive than the HDF search for
luminosities roughly greater than ${\rm Log}(L/L_{\odot})=-4.8$
despite the proper motion limit, while HDF is more sensitive
at lower luminosities.  (The exact location of the transition
depends on the age of the halo population.)  

We are now able to take both the LDM and the HDF surveys
into account by using the total effective volume searched, 
$V_{\rm tot}^{\rm eff}=V_{\rm LDM}^{\rm eff} + V_{\rm HDF}$. In the next section
we discuss how the combined survey
results, in comparison with model theoretical
luminosity functions, can constrain the age of the putative halo
population.
\bigskip
\centerline {\bf 4. RESULTS:}
\centerline {\bf COMPARISON OF MODEL LUMINOSITY FUNCTIONS WITH OBSERVATIONS}
\medskip
Here we compare the combined observational data from LDM and HDF
discussed in the previous section with theoretical 
model luminosity functions for white dwarfs.
The number of stars which should, on average, appear in a sample is
$$\nth=\int d\mbol \Phi(\mbol) V_{\rm tot}^{\rm eff}(\mbol) \, .
\eqno(4.1)$$
As discussed after Eq. (2.6), the luminosity functions
$\Phi(\mbol)$ are normalized to give a local halo white dwarf number
density $n_{\rm WD}$
appropriate to the choice of MACHO fraction $f$.
The $\dmax$ value for the HDF, 2 kpc, is substantially smaller
than the radius of the solar circle.  Thus, to first 
order, we can ignore details of the spatial distribution 
of white dwarfs and work with a single local density $n_{\rm WD}$.

As mentioned in the Introduction, we are employing the absence of dim
white dwarfs in the LDM and HDF surveys to determine lower bounds
on the age of a white dwarf halo population. If one assumes that the white
dwarfs are spatially uncorrelated, then in order to be consistent with
a nondetection at the 95\% confidence level, the number of dwarfs predicted
by a given model LF must be $n_{\rm th}\le3$. In this way, equation (4.1) can
be used for any model white dwarf luminosity function to restrict the 
local white dwarf halo density as a function of [{\bf 1}] the age of the 
white dwarfs, [{\bf 2}] the initial progenitor mass function, and [{\bf 3}]
the atmospheric composition. Variations in model luminosity functions
depend primarily on these three factors.

Figures 4 and 5 show the minimum age of a white dwarf halo
population, $t_{\rm halo}$, as a function of 
local white dwarf halo density $\rho_{\rm WD}$.  
In these figures,
regions above and to the left of a particular curve are excluded
at the 95\% confidence level. Excluded models thus predict at least 3 stars 
distributed through the effective volume of the LDM and HDF surveys.
Figure 4 shows our results when we use a log-normal IMF (following Adams
and Laughlin 1996) for the
main sequence progenitors, as motivated by star formation theory.
In Figure 5, on the other hand, we used delta function mass functions.
In both figures, we show the results obtained with both helium
and hydrogen atmospheres.  
In obtaining these figures, we have used the cooling curves
of Segretain \etal (1994).  In both figures, alternate cooling
models without C/O fractionation  would make all age limits younger by
2 Gyr.

Since we are investigating
the consequences of white dwarfs providing an explanation for the
MACHO data, as our lower limit for the MACHO fraction we take
$f> 0.1$ (i.e., MACHOs make up at least 10 percent of the halo).
We take 
the total local density of the halo to be
$10^{-2} {\rm M}_\odot {\rm pc}^{-3}$.
In figure 4, with the log-normal IMF, we then find that the youngest
allowed age for the white dwarfs is $\approx 13$ Gyr.
This minimum age
is obtained for the case of helium atmospheres with blackbody bolometric
corrections containing the ``safety factor'' (Eq. 3.3).  
Using the less conservative, but more likely blackbody BC
without the ``safety factor", we find that the minimum age rises to 
14 Gyr.  If the local halo white dwarf density were as 
high as $0.3\times10^{-2} {\rm M}_\odot /
{\rm pc}^3$, which is near the midpoint of the MACHO results
($\sim$30\% of the local halo density), 
the minimum age rises further to 15.5 Gyr.
Our minimum age is younger than what appeared to be allowed in
Chabrier {\it et al.}
(1996), who obtained 18 Gyr halo ages for $f$=0.25
as the approximate lower limits on the white dwarf age
(in the text of their paper, Chabrier \etal obtained a minimum
age of 16 Gyr for f=0.08).
The surprisingly young {\it minimum} age which we find
(the discrepancy with other authors) is a consequence of
{\it (i)} admitting MACHO fractions as low as $f$=0.1 to be marginally
consistent with the lensing result,
{\it (ii)} the proper motion limit in the Luyten Survey, {\it (iii)} 
conservative bolometric corrections, and {\it (iv)} the statistical process of
excluding models at the 95\% confidence level. Nevertheless, the general
result
of figure 4 is still suggestive of a white dwarf halo age commensurate with
the age of the globular clusters, as discussed below. 

Figure 5 shows how the limits on 
$\rho_{\rm WD}$ and $t_{\rm halo}$ depend on the 
form of the progenitor IMF. We have again plotted
the allowed halo age as a function of local halo white dwarf density,
and in this plot have assumed $\delta$
function IMFs for the white dwarf progenitors. As before, we
have considered the two cases of helium and hydrogen atmospheres.
Specifically, we have assumed that all white dwarf progenitor stars
in the sample had the same initial masses, ${\cal M} = 1.0, 1.5, 2.0, 3.0, 4.0,
5.0,$ and $6.0 M_{\odot}$.  Such extremely sharp
IMFs are of course unrealistic.  However, our results do illustrate that
the surveys are not particularly sensitive to higher mass, cooler white dwarfs,
while being extremely sensitive to WDs of initial mass 1--1.5 $M_{\odot}$.
Thus, an IMF weighted towards these low mass stars is the most
constrained.  For example, He atmosphere white dwarfs from
$1 M_\odot$ progenitors must be older than about 18 Gyr if they are to
contribute 10\% of the local halo density and still be consistent with the
combined survey results.
On the other hand, it is quite clear that an IMF weighted almost
entirely towards high mass initial stars is consistent with
a younger Halo. As seen in Fig. 5, if all the main sequence
progenitors of the white dwarfs initially were 6 $M_\odot$
stars with Helium atmospheres, then the halo could be less
than 12 Gyr old.  Discussion of these age limits with
possible cosmological implications is presented shortly
in the following section.

Both figures emphasize that white dwarfs with hydrogen atmospheres (or at least with atmospheres whose bolometric corrections are similar to those we have adopted)
cannot provide more than a few percent of the mass of the halo.
Here we have taken 18 Gyr as the upper limit to the age of the galactic
halo, since this is the oldest possible age of globular 
clusters (Chaboyer \etal 1996).  As discussed earlier in the text,
hydrogen atmosphere white dwarfs are much brighter than
those with helium atmospheres, and are thus far more constrained.
Hence our results suggest that if
white dwarfs are indeed present at the level suggested by the MACHO
survey, the majority must have helium atmospheres.
To reiterate, the absolute lower limit on the age of the white
dwarfs is provided by the pure Helium atmosphere white dwarfs;
these lower limits will potentially be revised upward in the future
as uncertainties are reduced.

{\it A note on uncertainties}:  Uncertainties in the white dwarf ages
come from a number of factors.  We have tried to
be very careful to take into account the
three factors that lead to the biggest uncertainties:
the nature of the white dwarf atmospheres, the bolometric
corrections, and the progenitor IMF.  As we have seen,
H atmospheres in the white dwarfs lead to ages that are larger
than for He atmospheres
by more than 8 Gyr.  The IMF leads to at least several Gyr
of uncertainty, as indicated by the $\delta$ function IMF exercise
shown in Fig. 5.  For He atmospheres, we have compared blackbody
BCs with and without the ``safety factor" of 0.8, and have
seen that this imparts about 1 Gyr of uncertainty.  Here we
have been extremely conservative in allowing such a large
safety factor, which leads to ages 1 Gyr younger. 

As mentioned previously,
there may be uncertainty associated
with the white dwarf cooling theory; this
uncertainty is $\sim 1$ Gyr. In obtaining the figures we have used
the models of Segretain \etal (1994); alternate models
without chemical fractionation lead to minimum ages that are 2 Gyr younger.
The initial/final mass relation imparts less than
0.5 Gyr of uncertainty.  This additional source
of uncertainty has therefore not been added into our final numbers.

\bigskip
\centerline {\bf 5. DISCUSSION}
\medskip
The second year MACHO events 
constitute a very provocative result. If
a lensing population of dim white dwarfs really accounts for a substantial
fraction of the missing halo mass, there are ramifications of
tremendous interest to cosmology, galactic evolution, star formation,
and white dwarf physics.  As a test of this white dwarf interpretation
of the MACHO data, we have considered the implications of {\it nondetections}
of white dwarfs in optical searches of
the Hubble Deep Field (Flynn, Gould, and Bahcall 1996)
as well as the ground based survey of Liebert, Dahn, and Monet (1988).
As long as MACHOs are indeed white dwarfs which make up
10\% of the local halo density, we are able to conclude
that (I) the halo must be at least 10 Gyr old 
(probably at least 14 Gyr) and (II) the halo
white dwarfs are likely to have He dominated atmospheres.

{\bf 5.1 Cosmological Implications:} 

These ages are consistent with the age of the oldest globular clusters in
the Galaxy, 11.6-18.1 Gyr (95\% confidence, Chaboyer \etal 1996).
Observations of halo white dwarfs provide an important and largely independent
estimate of the age of the galactic halo, and hence a lower bound of
the age of the universe.
Such an old galaxy is difficult to reconcile with the Hubble age of
the universe.  For a Hubble constant $H_0=100h$ km sec$^{-1}$
Mpc$^{-1}$, in a flat, matter dominated universe without a
cosmological constant,
$$t_0=2/3 H_0^{-1}=6.5h^{-1}{\rm Gyr}$$
For an age limit $t_0>12$ Gyr, we have $h<0.55$.  For a stronger
age limit of 14 Gyr, we have $h<0.46.$

There are several competing
observations of the Hubble Constant.
Madore \etal (1996) measured the distance to Cepheids in the Fornax
Cluster using
the Hubble Space Telescope and derived $0.63<h<0.97$.  Measurements of
Supernovae 1A's have set the Hubble constant to be $0.6<h<0.74$ (Riess
\etal 1995).  On the other hand, Tammann \etal (1997) have also measured
distances to Cepheids in the Virgo cluster with the HST and have
derived a value $0.45<h<0.65$.  Measurements of time delays
in a gravitationally lensed quasar yield estimates of $h$ (for $\Omega=1$)
of 0.63$\pm$0.12 (Kundic \etal 1997).

Only the low range of these values of $h$ is consistent with an
age limit of 12 Gyr.  If $h>0.6$,
and if our estimate of the local halo white dwarf density is correct,
we would be forced to abandon the standard, flat, matter dominated
cosmology.  One possible alternate cosmology is an open universe, in
which the factor of 2/3 in equation 5.1 is closer to unity.
For a lower limit of 12 Gyr, 
one requires $h < 0.67$ for $\Omega = 0.3$,
and $h < 0.83$ for $\Omega =0$. These higher values of $h$ are
somewhat easier to reconcile with all the data.
A cosmological constant might also imply an older universe.

{\bf 5.2 Other Implications of and Problems with a White Dwarf Halo:} 

{\it Chemical Evolution:} A large halo white dwarf population would
affect the chemical evolution of the Milky Way.  The primary difficulty
with the hypothesis that the lensing MACHOs are white dwarfs appears to
lie with the copious quantities of enriched gas which the stars that
formed in an early wave of star formation would later have deposited
into the interstellar medium.  Stars that become white dwarfs change the
chemical composition of the interstellar medium.  A 2$M_\odot$ star will
become a 0.6$M_\odot$ white dwarf, spitting out 70\% of its mass in the
form of hot, Helium rich, Carbon and Nitrogen rich, deuterium depleted
gas.  This gas has long been used to place limits on white dwarfs as
baryonic dark matter candidates (see Carr 1994 for a review; also
see Smecker and Wyse 1991).  In
addition, a recent chemical evolution study carried out by Gibson \&
Mould (1997) shows that the nucleosynthetic yield of a ``Population
III'' halo white dwarf precursor population would produce a CNO
abundance ratio of [C,N/O]$\approx$+0.5, which stands in stark
disagreement with the [C,N/O]$\approx$-0.5 ratio observed in current day
halo stars (Timmes \etal 1995). 

It appears that a large quantity of the
enriching gas would have to have been driven from the halo in order to
explain current population II abundance ratios; i.e., the ``closed box"
model of the Galaxy would have to be abandoned.  Such scenarios for
ejecting metals from the Galaxy have been outlined by Scully \etal
(1996) and Fields, Matthews \& Schramm (1996).  It is also intriguing to
note that studies of chemical evolution in rich clusters of galaxies
suggest that these galaxies had an early burst of star formation that
produced high mass stars (\eg Elbaz, Arnaud \& Vangioni-Flam 1995).
According to these models, supernovae from these high mass stars
generated a galactic wind which blew most of the metals out into the
intergalactic medium.  A future paper (Graff, Fields, Vangioni-Flam \&
Freese, in preparation) will deal with this issue. 

{\it Ionizing the Intergalactic Medium:} The light from all the white
dwarf progenitors would also help ionize the intergalactic medium and
could thus contribute to an understanding of the Gunn-Peterson test
(Gunn \& Peterson, 1965), which showed a dearth of background 
neutral hydrogen.
Miralda-Escude and Ostriker (1990) and Giroux and Shapiro (1996)
showed that the light from even the disk population of stars can make
a significant contribution to the ionization of the intergalactic
medium.  The additional light from a large, high initial mass halo
population of white dwarf progenitors would make a much greater
contribution to the ionization than the disk population alone.

{\it Light emitted by High Redshift Galaxies:} The large population of
halo white dwarfs would have been very bright when they were main
sequence stars and should dominate the light emitted by high redshift
galaxies.  It may be that the detected excess population of distant
blue galaxies (Tyson 1988) is due to the light emitted during the main
sequence phase of what later became white dwarf MACHOs.  Charlot and
Silk (1995) examined the evolution of light emitted by white dwarf
halos and found that for most of their models, they predicted more dim
galaxies than are observed. New and higher quality data on the
luminosity functions of high redshift galaxies is pouring in.  For
example, Steidel \etal (1996) found a population of blue star-forming
galaxies at redshift $z=3$.  The contribution of the white dwarf
progenitors to the light of high redshift galaxies will remain an
interesting subject.  A future paper (Graff, Devriendt, Charlot \&
Guiderdoni, in preparation) will deal with this subject.

{\bf 5.3 Proposed Improvements:}
Our main goal in this paper has been to re-evaluate the age limits on
the halo population. A significant new aspect of our analysis
is the specific recognition of the importance of the velocity bias in
the LDM survey. The result of this bias is that the LDM survey places
considerably less strict limits on 
the number of dim white dwarfs than previously believed. Indeed, the
effect of the bias is so important that it renders the LDM survey
{\it less} sensitive than the HDF for luminosities below
$\lglsun=-4.8$.
At white dwarf luminosities of $\lglsun=-5.0$
the bias reduces the effective LDM search volume by nearly a factor of 10.

There is still a great deal of room for progress in the art of limiting the 
density of Halo white dwarfs, both theoretically and observationally. 
Due to its proper-motion biases, the Luyten survey used by LDM was 
clearly not an ideal
place to search for a population of {\it halo} white dwarfs.  However, a
new survey could quite feasibly be done with ground based telescopes.  Modern
CCD's are sensitive in the infra-red where Luyten's survey was
not, and one could easily search for stars with proper motions greater than
$\mu=2.5 ''/yr$.  Although such a search would not necessarily go as deep
as the HDF, it could potentially cover $4\pi$str$/4.4$arcmin$^2=3.4\times 
10^7$ times as much solid angle.  We mention here three
such searches that are ongoing or proposed: Three square degrees down to 
$I=22$ were observed by Lidman and Silk and are now 
being processed.  200 square degrees down to $I=21$ are presently
being observed by the EROS team.
A very promising survey has been proposed as a project to be
undertaken with the forthcoming SUBARU 8 meter telescope (Alcock,
personal communication) which would observe a large area of the sky 
in the $V$ and $I$ bands down to limiting magnitude of 29 or 30.

Beyond possibly explaining the lensing results, the optical identification
of exceedingly dim white dwarfs in a deep proper motion survey of this type 
would be
of great importance. From a purely astronomical point of view, these would be 
the oldest, coolest white dwarfs yet seen.
They would allow us to test
white dwarf cooling theory into the Debye crystallization regime and 
calibrate models of extremely cool white dwarf atmospheres.

While there has been recent progress in extending white dwarf cooling curves
down to very low temperatures, the computation of very cool, frequency
dependent white dwarf atmosphere models has lagged considerably. Now that
the microlensing searches suggest that a hitherto undetected population of very
cool white dwarfs is lurking in the Galactic halo, new atmosphere models
are needed for temperatures cooler than 4000K.  
As can be seen by comparing the black body curve 
and our conservative He atmosphere curve in figure 4, uncertainty in the 
bolometric correction adds about 1 Gyr of uncertainty in the 
interpreted age of a population of white dwarfs.  A project 
is currently underway by M\'{e}ra and Allard to 
generate these cool model white dwarf atmospheres.
The peak emission of
these cool white dwarfs is in the infrared; the exact 
peak wavelength of emission depends on the white dwarf age and properties,
and will be easier to predict once theoretical advances occur
in understanding white dwarf atmospheres and bolometric corrections.

{\it Binaries:}
Another possibility for refining the theoretical model luminosity
functions is to take binary systems into account:  the white dwarfs
may have binary companions.  Roughly half of
Population I stars are in binaries; thus it is plausible
that some fraction of the halo white dwarf stars
may also be in binary systems. [Of course the halo white dwarf
stars we are considering here come from a very different IMF
than do the Pop I stars; thus the binary distribution could also be
different].  
In order to examine how much binaries can change our final result,
we have considered the extreme case in which {\it all} white dwarfs
have unresolved binary companions, and both companions are identical. 
Each unresolved binary acts like a ``star" which is twice
as bright as the luminosity of the component stars.
In a pure magnitude limited sample, doubling the luminosity increases the volume
searched by a factor of $2^{3/4}$.  Due to the velocity biases of the LDM
survey, the volume actually increases somewhat faster in that survey;
it roughly increases by a factor of 4.  Since these binary systems
are theoretically expected to be brighter at a given age, they must
be older to have escaped detection.  On the other hand, each
binary system has twice the mass of a single star, so that half as
many are needed to account for the local halo white dwarf mass density.
Thus a MACHO density that is 10\% of the Halo
could be accounted for by a density of white dwarf binary systems that
is only 5\% of the halo.  When we redid our calculations 
using this extreme binary model, we found that the lower 
limits on ages increased by about 0.6 Gyr.

Some of these binary systems may be in very close orbits.
Such a population may cause type 1a Supernovae.  Smecker \& Wyse
(1991) estimated the SN1a rate from a halo full of binary white dwarfs
and concluded that their models predicted 100s of times too many
SN1a's.  
While theirs is a severe constraint, their
assumptions about mass loss, orbital radius, and binary fraction may
not apply to a zero metallicity population.  Gravitational radiation
from a population of close binaries would be detectable with the
proposed OMEGA experiment.

{\bf 5.4 Conclusion}

Our work was motivated by microlensing measurements which suggest that
MACHOs have masses consistent with their being white dwarfs.
Of course it is always possible that the MACHOS that are being
seen are in fact different objects entirely; however, in this
paper we examined the possibility of a substantial white dwarf halo.
In conclusion, we have found an age constraint
on white dwarfs that are candidates for explaining the microlensing
events.  If white dwarfs comprise at least 10\% of the halo of the
galaxy, then with extreme parameters the minimum age we obtain
for the white dwarf population is 10 Gyr.
With more likely parameters and up-to-date cooling curves,
the white dwarfs must be at least 14 Gyr old to be
marginally consistent with the joint 
results of the LDM, HDF, and MACHO surveys. We used
recent theoretical cooling curves of Segretain \etal (1994)
and considered
a variety of progenitor IMFs, bolometric corrections and atmospheric
compositions to come to this conclusion.  Analysis of
the effects of the velocity bias in the LDM
survey also played a significant role in our results.
Potentially the most serious problem with a white dwarf halo lies 
with the enriched gas the progenitors would have produced.

\bigskip 
\bigskip 
\centerline{Acknowledgements} 

We would like to thank Fred Adams, Gilles Chabrier, Chris Lidman, Didier Saumon, 
and Matt Wood for
assistance, and Conrad Dahn for useful discussions concerning the 
proper motion limits in the Luyten catalogue.  Gilles Chabrier graciously
provided us with our white dwarf cooling curves. We acknowledge support
from NSF grant PHY-9406745, and from the University of Michigan Physics
Department.
\medskip

\np 
\centerline{\bf REFERENCES}
\medskip

\rfnce{
Adams, F. C. \& Laughlin, G. P. 1996, ApJ, 468, 586}
\rfnce{
Adams, F. C. \& Fatuzzo, M. 1996, ApJ, 464, 256}
\rfnce{
Alcock, C. \etal 1997, ApJ submitted}
\rfnce{
Allard, F. 1990 Ph.D. thesis, Univ. Heidelberg}
\rfnce{
Ansari {\it et al.} 1996 AA, 314, 94}
\rfnce{
Bahcall, J. \& Soneira, R. M. 1980 ApJS, 44, 73}
\rfnce{
Bergeron, P., Ruiz, M. T. \& Leggett, S. K. 1997, ApJS, 108, 339}
\rfnce{
Bergeron, P., Saumon, D., \& Wesemael, F. 1995 ApJ, 443, 774}
\rfnce{
Carr, B., 1994, ARAA, 531, 32}
\rfnce{
Chaboyer, B., Demarque, P., Kernan, P. J., Krauss, L.M. 1996,
Science, 271, 957}
\rfnce{
Chabrier, G., Segretain, L., \& M\'{e}ra, D. 1996, ApJ, 468, L21}
\rfnce{
Charlot, S. \& Silk, J. 1995, ApJ, 445, 124}
\rfnce{
Dahn, C. C., Liebert, J, Harris, H. C. \& Guetter, H. H. 1995,
in ``The Bottom of the Main Sequence and Beyond''
ed. C. G. Tinney (Springer-Verlag, Heidelberg), 239}
\rfnce{
Elson, R. A. W., Santiago, B. X. \& Gilmore, G. F. 1996, NewA 1, 1}
\rfnce{
Fields, B., Matthews, G. \& Schramm, D. S. 1996, Preprint, {\tt
astro-ph \# 9604095}}
\rfnce{
Flynn, C., Gould, A., \& Bahcall, J. 1996, ApJ, 466, L55}
\rfnce{
Fontaine, G. \& Wesemael, F. 1987, in IAU Colloq. 95, The Second Conference
on Faint Blue Stars, ed. A. G. D. Philip, D. S. Hayes \& J. Liebert
(Schenectady: L. Davis Press), 319}
\rfnce{
Gates, E. \& Turner, M. S. 1994, PRL, 72, 2520}
\rfnce{
Gibson, B.K. \& Mould, J.R. 1997, ApJ, In Press}
\rfnce{
Giroux, M. L. \& Shapiro, P. R. 102, 191}
\rfnce{
Graff, D. S. \& Freese, K. 1996, ApJ, 456, L49}
\rfnce{
----- 1996, ApJ, 467, L65}
\rfnce{
Gunn, J. E. \& Peterson, B. A. 1965, ApJ, 142, 1633}
\rfnce{
Iben, I. Jr., \& Laughlin, G. 1989, ApJ, 341, 312}
\rfnce{
Iben, I. Jr. \& MacDonald, J. 1986 ,ApJ, 301, 164}
\rfnce{
Iben, I. Jr. \& Tutukov, A.V., 1984, ApJ, 282, 615}
\rfnce{
Kawaler, S. D. 1996, ApJ, 467, L61}
\rfnce{
Kundic, T. \etal 1997 preprint, {\tt astro-ph 9610162}}
\rfnce{
Liebert, J., Dahn, C. C. \& Monet, D. G. 1988, ApJ, 332, 891 (LDM)}
\rfnce{
Liebert, J., Dahn, C. C., Gresham, M. \& Strittmayer, P. A. 1979, ApJ,
233, 226}
\rfnce{
Luyten, W. J. 1979, LHS Catalogue (Minneapolis: University of
Minnesota Press)}
\rfnce{
Madore, B. F., \etal 1996, BAAS, 189, 108}
\rfnce{
Miralda-Escud\'{e}, J. \& Ostriker, J. P. 1990, ApJ, 350, 1}
\rfnce{
Oswalt, T. D., Smith, J. A., Wood, M. A. \& Hintzen, P. 1996, Nature,
382, 692}
\rfnce{
Reid, I. N., Yan, L., Majewski, S.,Thompson, I. \& Smail, I
1996, AJ, 112, 1472}
\rfnce{
Riess, A.G., Press, W. H. \& Kirshner, R. P. 1995, ApJ, 445, L91}
\rfnce{
Saumon, D., Bergeron, P., Lunine, J. I., Hubbard, W. B. \& Burrows,
A. 1994, ApJ, 424, 333}
\rfnce{
Segretain, L., Chabrier, G., Hernanz, M., Garcia-Berro, E.,
Isern, J. \& Mochovitch, R. 1994, ApJ, 434, 641}
\rfnce{
Sion, E. 1986, PASP, 98, 821}
\rfnce{
Smecker, T. A. \& Wyse, R. F. G. 1991, ApJ, 372, 448}
\rfnce{
Steidel C. C., Gialvalisco, M., Pettini, M., Dickinson, M., \&
Adelberger, K. L. 1996, ApJ, 462, L17}
\rfnce{
Tamanaha, C., Silk, J., Wood, M. A. \& Winget, D. E. 1990, ApJ, 358, 164}
\rfnce{
Tammann {\it et al.} 1997, Preprint {\tt astro-ph 9603076}}
\rfnce{
Timmes, F.X., Woosley, S.E. \& Weaver, T.A. 1995, ApJS, 98, 617}
\rfnce{
Turner, M. S., Gates, E. \& Gyuk, G. 1996, Preprint {\tt astro-ph 9601168}}
\rfnce{
Tyson, J. A. 1988 AJ, 96, 1}
\rfnce{
Williams, R. E. \etal 1997, Preprint, {\tt astro-ph 9607174}}
\rfnce{
Winget, D. E., Hanson, C. J., Liebert, J., Van Horn, H.
M., Fontain, G., Nather, E., Kepler, S. O., \& Lamb, D. Q. 1987, ApJ,
315, L77}
\rfnce{
Wood, M. 1992, ApJ, 386, 539
\rfnce{
Wood, M.  1994, in ``White Dwarfs'' D. Koestler \& K. Werner (Eds.)
Springer-Verlag, 41}

\vfill\eject
\centerline{\bf FIGURE CAPTIONS}

\vskip 0.07 in\noindent
Figure 1. ---
The luminosity of white dwarfs as a function of initial mass. From
top to bottom, the curves show the luminosity of Helium atmosphere
white dwarfs which were formed as {\it main sequence} stars 10--18
Gyr ago. The horizontal line shows the bolometric luminosities of
the dimmest white dwarfs observed by LDM if one assumes a black body
bolometric correction.

\vskip 0.07 in\noindent 
Figure 2. --- 
>From bottom to top, the three sets of curves indicate bolometric
corrections for the $V$, $R$, and $I$ bands.  The solid lines are black
body bolometric corrections.
The crosses and triangles are the calculated white dwarf
bolometric corrections from Bergeron, Saumon \& Wesemael (1995) for
Helium and Hydrogen atmospheres, respectively.  Note that these
bolometric corrections do not extend below $\teff=4000K$.  The
pentagons connected by long-dashes are the zero-metallicity red-dwarf
calculations of Saumon \etal (1994), which we have adopted as an
underestimate of the Hydrogen atmospheres (triangles).  The short-dash
curves show the shifted black-body bolometric corrections which we have adopted 
as an underestimate of the Helium atmosphere bolometric corrections (crosses).

\vskip 0.07 in\noindent
Figure  3. ---
Effective volumes probed in the surveys under consideration. 
Hydrogen atmosphere bolometric corrections have been used in the left panel,
whereas helium atmosphere bolometric corrections have been used in the right
panel. Both panels show the effective volumes probed by (from bottom up)
the HDF, the LDM data (culled from Luyten's survey),
and the effective volume LDM would have sampled had Luyten's survey contained
no proper motion cutoffs. 
Bolometric corrections are calculated for stars aged 13 Gyr.  There is
a small feature at 4000K where we switch from the Bergeron,
Saumon and Wesemael model BCs to estimated BCs.

\vskip 0.07 in\noindent
Figure  4. --- 
Excluded Halo ages as a function of local white dwarf halo density, under
the assumption of Adams \& Laughlin's log-normal IMF for the white
dwarf progenitors, with $m_C=2.3 M_{\odot}$ and $\sigbar=0.44$.
The shaded region of parameter space is excluded
with 95\% confidence if the white dwarfs have shifted blackbody
bolometric corrections appropriate to a Helium atmosphere. The other
curves show the 95\% confidence contours under the assumptions that
(1) the Helium atmospheres have unshifted blackbody corrections,
and (2) the stars have Hydrogen-atmosphere bolometric corrections
as estimated in the text.
The region above the horizontal solid line indicates a MACHO fraction
that is at least $10^{-3} M_\odot/{\rm pc}^3$ (i.e., at least 10\% of
a total local halo density that is $10^{-2} M_\odot/{\rm pc}^3$).
This figure has been obtained using white dwarf cooling theory
of Segretain \etal (1994).  A cooling theory without chemical
fractionation would result in age limits that are 2 Gyr younger than what is
presented in this plot. 

\vskip 0.07 in\noindent
Figure  5. ---  
Excluded Halo ages as a function of local white dwarf 
Halo density under the assumption of
idealized $\delta$ function IMFs for the white dwarf progenitors. 
In all cases, the parameter space above the lines is excluded
with 95\% confidence.  The solid line labeled 1.5
near the center of the plot refers to the case where
all halo white dwarfs have Helium
atmospheres and emerged from stars of initial mass
1.5 $M_{\odot}$. As one progresses up and to the left
on the plot, similar lines are shown constraining Helium atmosphere
white dwarfs arising from progenitor masses of 2,3,4,5, and
6 $M_{\odot}$. Similarly, the bundle of lines at the 
lower right shows 95\% confidence limit
contours computed for Hydrogen atmospheres.
A curve representing the 95\% confidence limit for 
Helium stars with initial mass 1.0 $M_{\odot}$ has been left out 
for clarity, as it closely mimics the 5  $M_{\odot}$ Hydrogen curve.
Again, the region above the horizontal solid line indicates a MACHO fraction
that is at least $10^{-3} M_\odot/{\rm pc}^3$ (i.e., at least 10\% of
a total local halo density that is $10^{-2} M_\odot/{\rm pc}^3$).
This figure has been obtained using white dwarf cooling theory
of Segretain \etal (1994).
A cooling theory without chemical
fractionation would result in age limits that are 2 Gyr younger than what is
presented in this plot. 

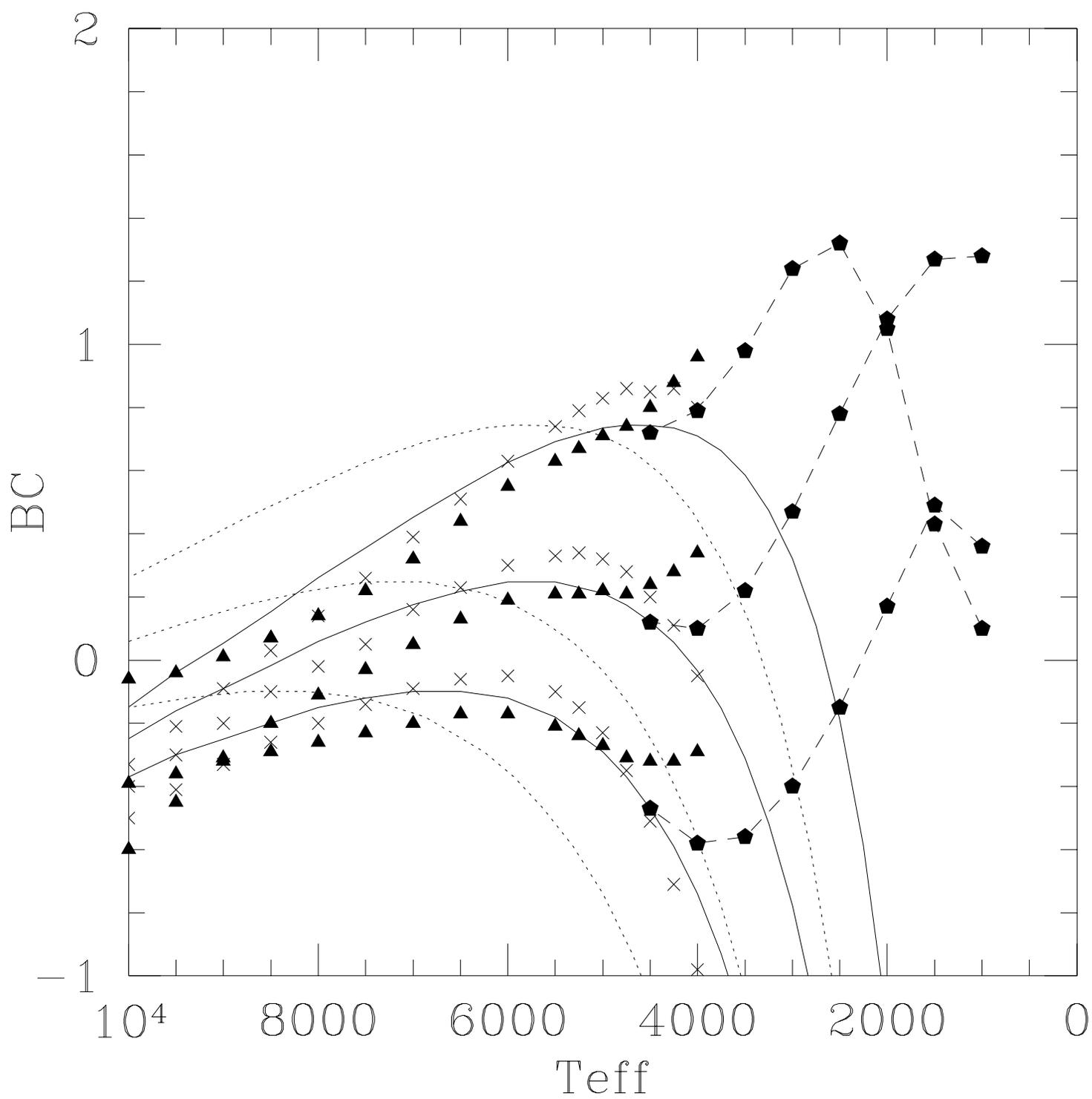
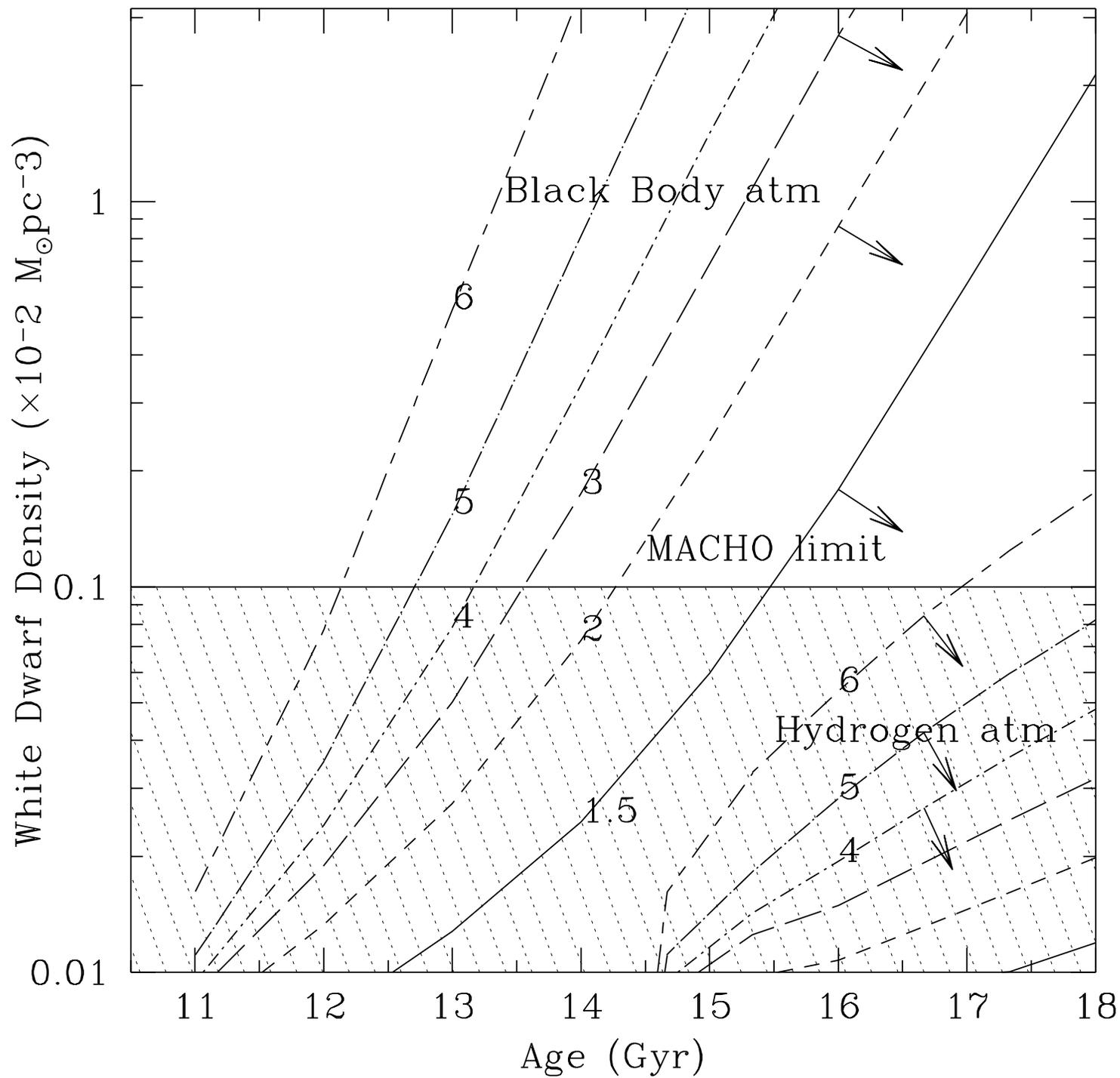
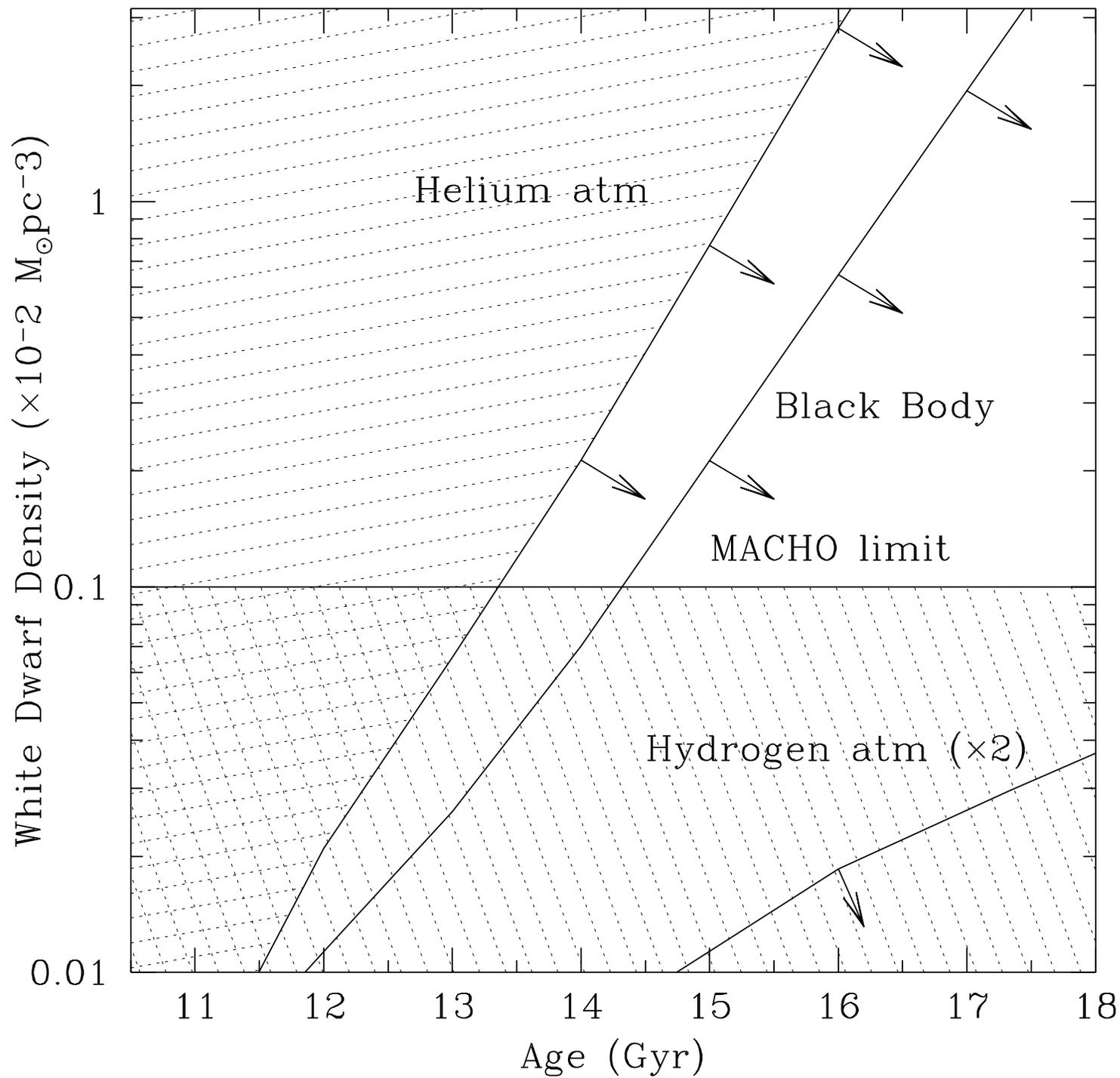
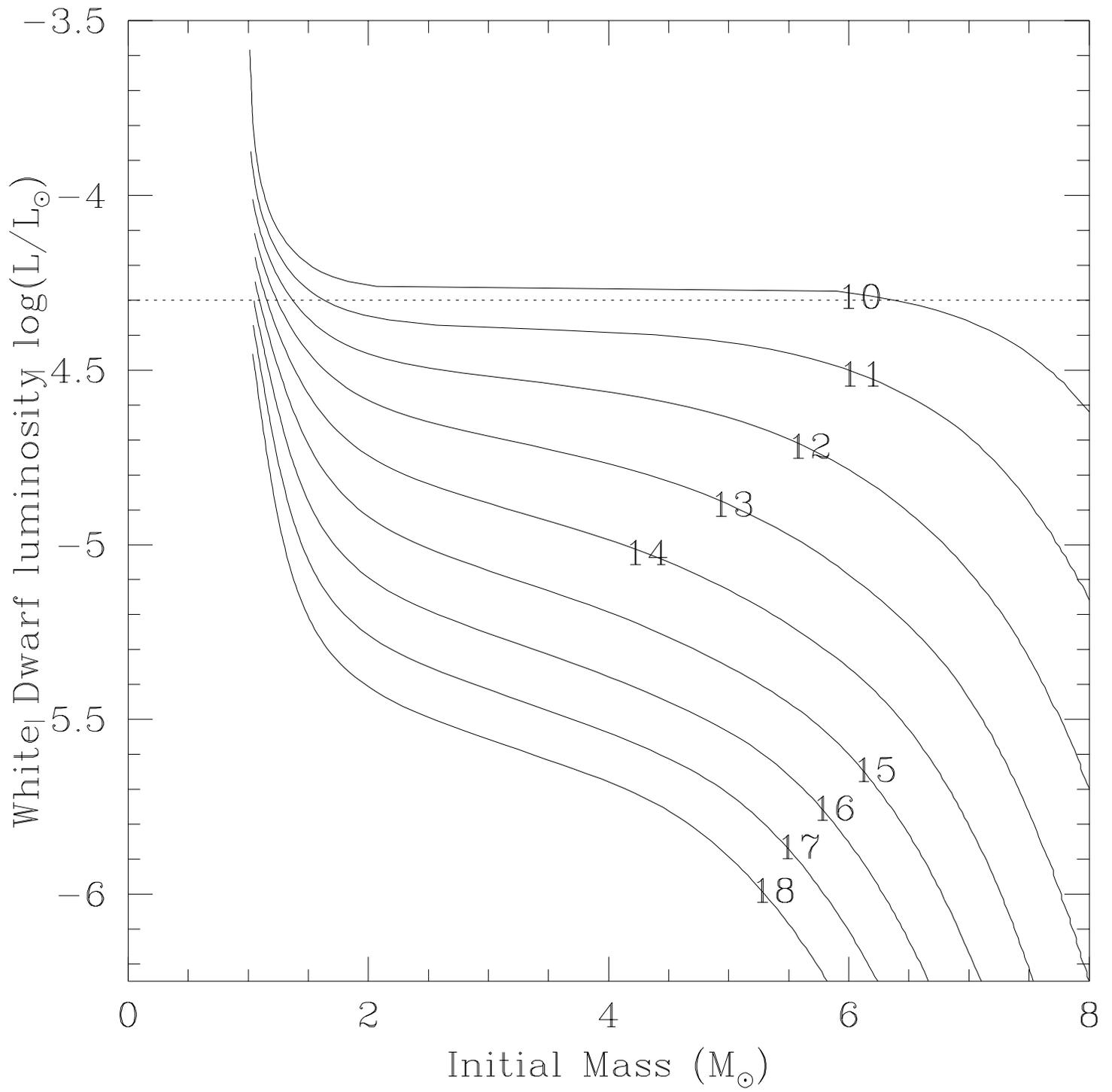
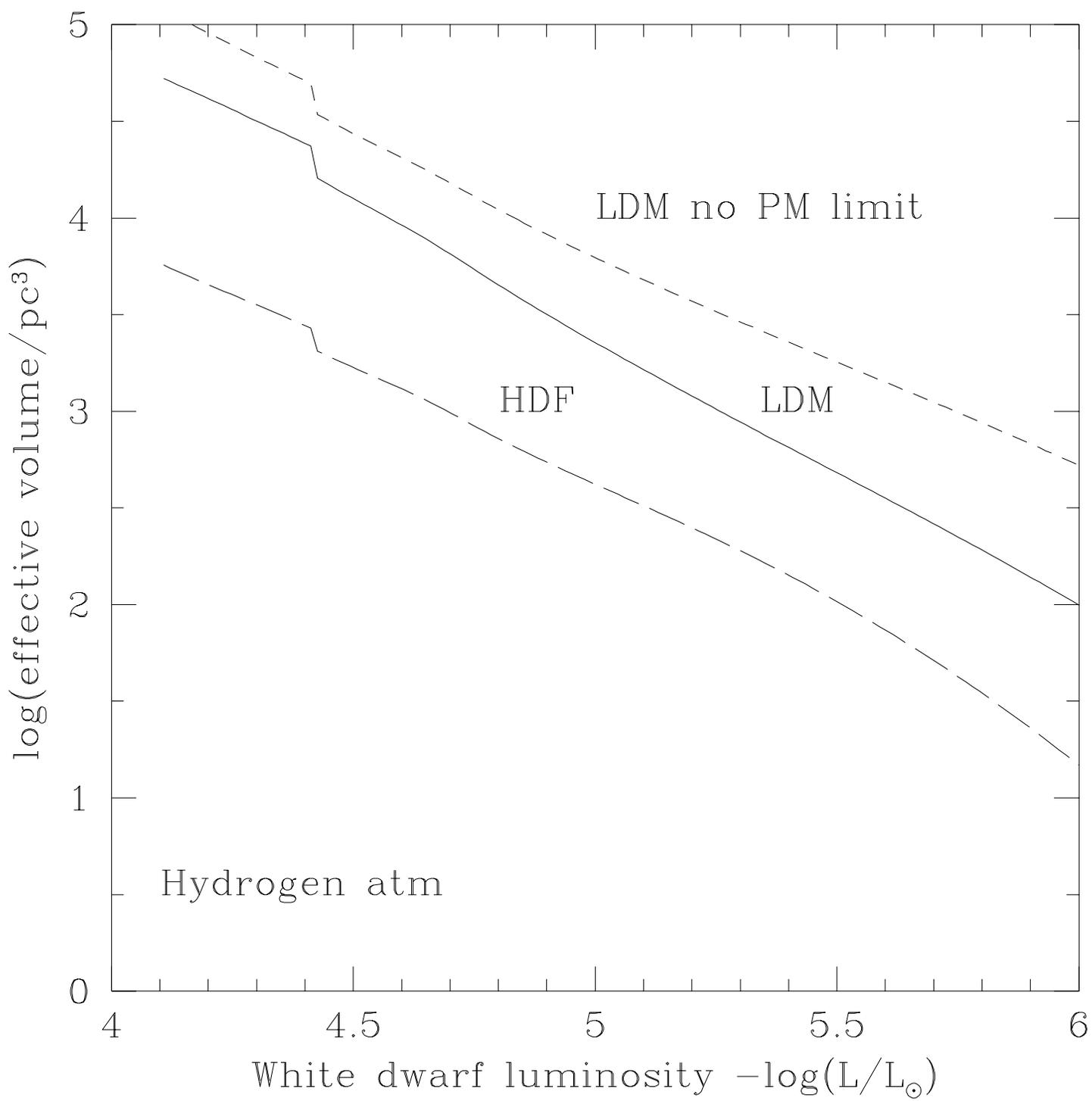
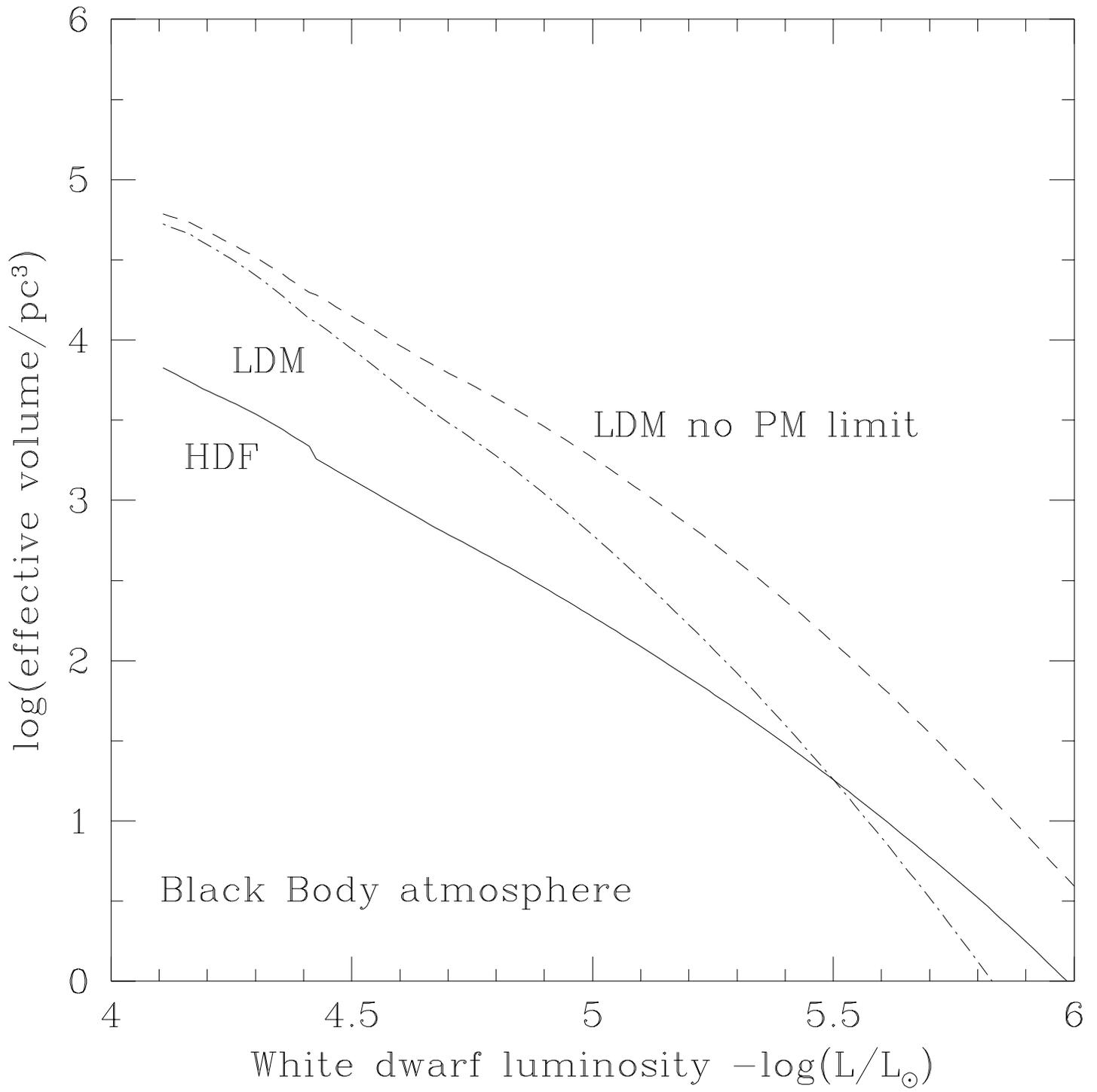
\bye